\documentclass[10pt,german,english,twoside]{article}
\usepackage[latin1]{inputenc}
\usepackage{eurosym}
\usepackage{german}
\usepackage[hscale=0.75,vcentering=true,vscale=.8]{geometry}
\usepackage{amssymb}
\usepackage{amsmath}
\usepackage{fancyhdr}
\usepackage[sl]{caption2}
\usepackage[sort&compress,square,comma,numbers]{natbib}
\usepackage{wasysym}
\usepackage{tabularx}
\usepackage{arydshln}
\usepackage[]{graphicx}
\usepackage{url}
\usepackage{epsf}
\usepackage{calc}
\usepackage{setspace}
\usepackage[german]{babel}
\usepackage{subfigure}
\usepackage{makeidx}
\usepackage[]{color}
\usepackage{soul}
\usepackage[collision]{chemsym}
\usepackage[ colorlinks, citecolor=black,linkcolor=black, urlcolor=black, pagebackref=false, bookmarks=true, bookmarksopen=true, bookmarksnumbered=true]{hyperref}
\usepackage{rotating}
\usepackage{soul}

\begin{document}
\begin{center}
\huge
Advanced superconducting circuits and devices\\
\Large
To be published in \emph{Handbook of Applied Superconductivity}, Wiley-VCH\\
Martin Weides and Hannes Rotzinger\\
Physikalisches Institut, Karlsruher Institut f\"{u}r Technologie, Germany
\end{center}
\normalsize
\tableofcontents \par

The unique properties of superconducting devices like ultra low power consumption, long coherence, good scalability, smooth integration with state of the art electronics and in combination with the Josephson effect outstanding magnetic field sensitivity, suits them well for a wide range of applications.
Questions in fundamental and applied disciplines, like material characterization, particle detectors,  ultra-low loss microwave components, medical imaging,  and potentially scalable quantum computers are commonly addressed.

Using integrated-circuit processing techniques, adapted from the semiconductor industry, complex micron or nanometer sized electronic circuits elements can be scaled up to large numbers.
Examples for such devices are for instance, SQUID or microwave resonators. SQUIDs (superconducting quantum interference devices) are high resolution magnetometers, sensitive enough to measure the extremely small magnetic fields originating, for example, from the human brain activity or single molecules.

Recent advances in thin film technology have made it possible to fabricate compact superconducting microwave resonators with a tremendous frequency purity. The developments towards such devices have been triggered by the requirement of storing the quantum information of a quantum bit as long as possible in an unperturbed way, since superconducting quantum circuits are one of the most promising systems to fulfill the DiVincenzo criteria \cite{DiVincenzoCriteria} for a scalable quantum computer.

Not only the quantum computing science is benefitting from this progress, but also other fields with functional micro and nano-structures in the ultra-low noise environment of low temperatures.
For instance, coupled cavity-qubit experiments in these circuits showed the validity of cavity quantum electrodynamics for microwave photons such as dressed states or quantum non-demolition readout. Parametric amplification of microwave signals was shown using a tunable resonator with an added noise temperature well below 1 Kelvin. Also very promising is the frequency domain multiplexing of many highly sensitive particle detectors each coupled to a superconducting resonator with a slightly different frequency. This approach allows for an energy sensitive particle camera with many pixels, being read-out by a single microwave line.

In the following we give an overview on superconducting circuits with new functionalities in the classical and quantum regime. We will start with a brief overview of novel devices based on nanoscale semiconducting Josephson barriers. Second, the fast-developing field of quantum computing is discussed with a focus on limitations, challenges, and opportunities. This is followed by a section on superconducting metamaterials and its applications at microwave frequencies in the classical and quantum regime. We conclude with an introduction to quantum phase slip  junctions and its potential being useful complementary to the Josephson junction.

\section{Field effect devices}

\subsection{Josephson field effect transistor}
\begin{figure}[]
\begin{center}
\includegraphics[width=5cm]{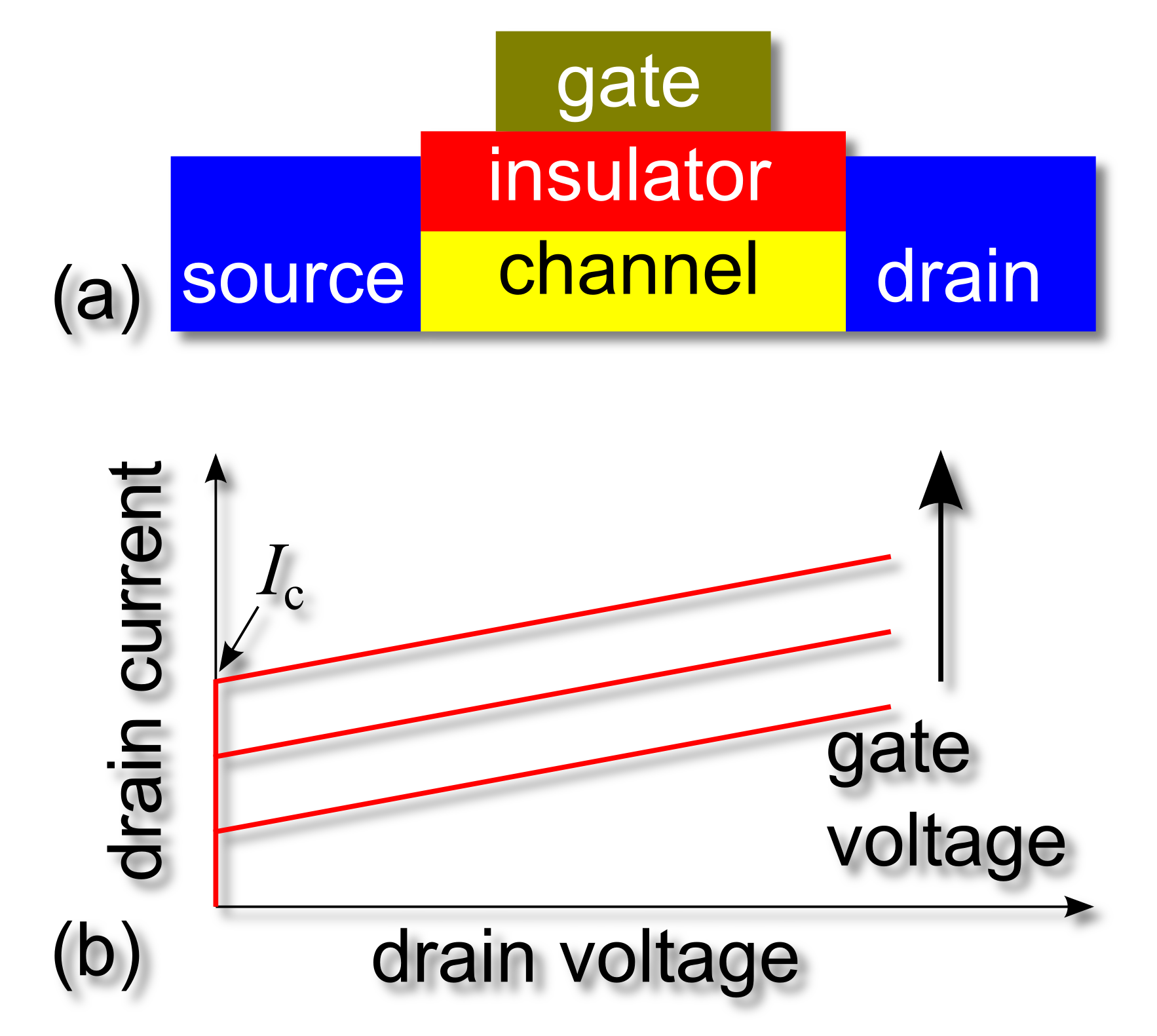}
\caption{\label{FET} {\bf a)} Schematics of a Josephson field effect transistor. The critical current $I_c$, and for larger source-drain currents the voltage between source and drain are modulated inside the channel by a voltage on the gate electrode. The insulator suppresses the leakage currents from the gate metal. {\bf b)}  Current-voltage characteristic of the source-drain channel as function of voltage bias on the gate electrode.}
\end{center}
\end{figure}

Field-effect superconducting devices use an applied electric field, and not the magnetic flux as normally done for conventional Josephson circuits, to control the coupling between the superconducting electrodes. Basic parameters, such as the critical current $I_c$, plasma frequency or characteristic voltage, see Fig. \ref{FET} can be in-situ modified.
The field-effect changes the Josephson coupling across a semiconducting material, for instance  with an applied voltage.
The effect is fundamentally different from the electrostatic tuning of the transport properties of small capacitance Josephson devices in the  Coulomb blockade / charge regime, which are  for instance used for superconducting  single electron tunneling devices or the charge qubit.

Josephson field-effect-transistors operate in the phase regime and are controlled by a gate electrode being galvanically decoupled from the junction by either a dielectric or a Schottky barrier \cite{ClarkJJFET_JAP80}.
$I_c$ and, for larger source-drain currents, the source-drain voltage are modulated as function of the gate voltage. Tunability was demonstrated for devices based on inversion layers, two-dimensional electron gases, nanowires and nanotubes. The superconductor/semiconductor interface transparency is critical, especially for silicon based semiconductors. They tend to form a Schottky barrier, unless very high doping concentrations (with reduced field-effect) are used. For voltage gain the output voltage across the source-drain channel must exceed the control voltage applied at the gate. Josephson field-effect transistors have been fabricated based on semiconducting interlayer \cite{Kleinsasser_JJSC_IEEE91}, channel \cite{BeckerPhysicsB95},  or wire \cite{Doh_Science05,Dam_Nature06,Xiang_Nature06,Frielinghaus_APL10,JarilloHerreroNat06} barriers. These systems have considerable potential for low-loss, fast digital elements as well as for analog applications such as cryogenic voltage amplifiers.

\subsection{NanoSQUIDs}
\begin{figure}[]
\begin{center}
\includegraphics[width=5cm]{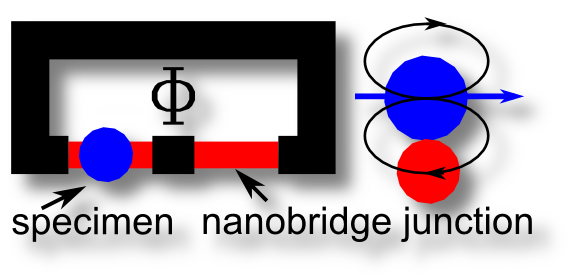}
\caption{\label{nanoSQUID}  {\bf Left:} A nanoSQUID with nanobridge Josephson junctions (red) having similar dimensions as the magnetic specimen (blue) under investigation.  {\bf Right:} Strong coupling of magnetic flux (black lines) from the specimen to the Josephson junction and SQUID loop (black) underneath significantly affects the sensed flux $\Phi$ through the DC SQUID loop. The nanoSQUID signal is detected either by transport or by microwave reflectometry.}
\end{center}
\end{figure}

Conventional SQUIDs (compare chapter 9) are made from Josephson tunnel junctions embedded in a closed loop with the junction's lateral dimensions being on the order of microns. Replacing the tunnel junctions with nano-scale Josephson junctions the noise performance can be enhanced \cite{VossAPL80} and single atomic spin measurement can be achieved \cite{Ketchen_APL84}. The detection of a single molecule's magnetic moment requires a substantial fraction of the particle's magnetic field to penetrate the SQUID loop. Choosing a nanoSQUID  \cite{FoleySUST09}, as depicted in Fig.~\ref{nanoSQUID}, the magnetic specimen can be placed above one of the Josephson junctions of similar dimensions. Thereby the coupling of magnetic flux to the loop is optimized.

The nanobridge junctions are typically formed by a superconducting constriction with a normal or semiconducting barrier made by thin films, nanowires or nanotubes. The nano-size dimensions are realized by patterning methods such as focused ion beam etching, electron beam or atomic force microscope lithography on lateral films. The interest in nanoSQUIDs goes well beyond the magnetic field detection. When using semiconductor barriers, e.g. carbon nanotubes \cite{Cleuziou_NatNanoTech06} or nanowires such as indium arsenide \cite{Dam_Nature06}, the junction behaves as a quantum dot with well-separated electronic levels due to the exciton confinement in all three spatial dimensions. The position of the quantum levels can be tuned with an applied gate voltage similar to the Josephson field-effect barriers described in the previous section.
Thereby a gate-controlled transition from the normal ($0$-type) to the $\pi$-Josephson junction can be observed (see chapter 1) depending on the electron parity in the constriction. These junctions have a Josephson phase $\phi=\pi$ in the ground state when no external current or magnetic field is applied, and a  negative critical current $I_c$ with the first Josephson relation being modified to $I_s = -|I_c|\sin(\phi) = |I_c|\sin(\phi+\pi)$.

\subsection{Majorana fermions and topological qubits}

Another field of significant interest are semiconducting nanowires controlled by local gates and proximitized by a superconductor. They receive considerable attention in the quest for Majorana fermions and topological qubits. Majorana fermions are fermionic quasiparticle excitations which are their own anti-particles \cite{Leijnse_Majorana12}.

\begin{figure}[]
\begin{center}
\includegraphics[width=5cm]{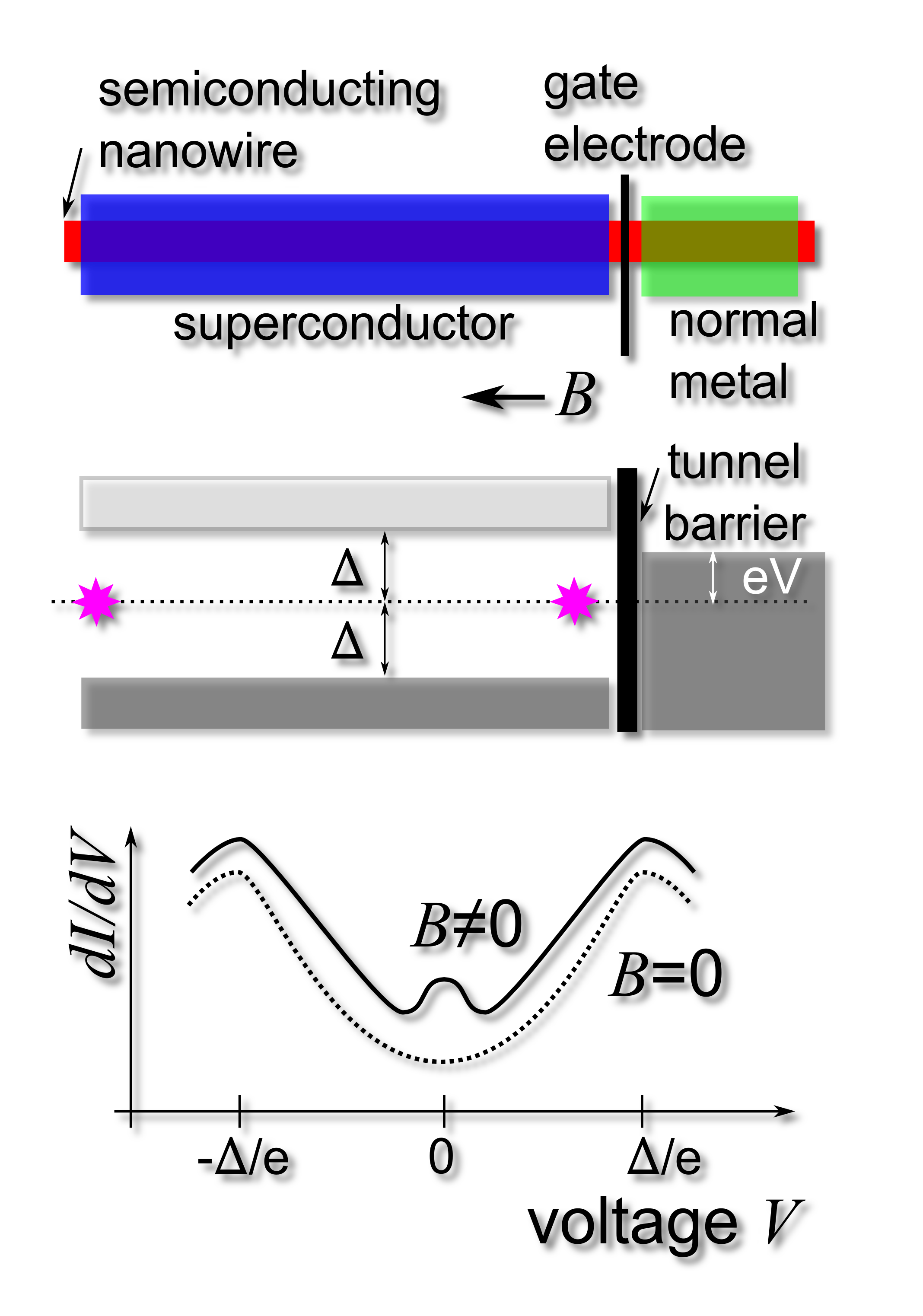}
\caption{\label{Majorana}  {\bf Top:} The nanowire with strong spin-orbit coupling is proximity coupled to a bulk superconductor and a normal metal. A gate voltage establishes a tunneling barrier at the interface between the metallic and superconducting parts of the wire. An applied magnetic field $B$ parallel to the wire induces two Majorana fermions (depicted as stars) as quasiparticles on both interfaces. {\bf Middle:} For bias voltages $V$ smaller than $\Delta/e$ electrons from the normal metal cannot tunnel into the wire. In case a Majorana state is located near that tunnel barrier, electron tunneling into the wire's Fermi level is allowed. {\bf Bottom:} The differential conductance $dI/dV$ across normal-metal and superconducting electrodes as a function of applied bias voltage $V$ shows a local maximum at zero voltage indicating the emergence of a Majorana fermion inside the gap and next to the tunnel barrier.}
\end{center}
\end{figure}

Majorana fermions are expected to be located in conductive nanowires contacted with superconductors. In the proximitized region they may emerge as non-fundamental quasiparticles at zero-energy. Since the Fermi level is in the middle of the superconducting gap, these midgap states can be realized by conventional thin films technology, nanowires of semiconductors \cite{Alicea_PRB10,Sau_PRL10} or metals \cite{Potter_PRB12} with a pronounced spin-orbit coupling. An external magnetic field $B$ is applied to make the band appear spinless. For certain electron densities, the combination of the induced superconducting gap, the strong spin-orbit coupling in the nanowire and the magnetic field open an unconventional superconducting state with Majorana bound states at its ends, see Fig. \ref{Majorana}. Recently, its signature was reported based on electrical measurements on an indium antimonide nanowire-superconductor hybrid structure \cite{Mourik_Science12} where gate electrodes modified the electronic structure of the nanowire exposed to a magnetic field. The conductance close to zero voltage is enhanced by the presence of Majorana states next to the tunnel barrier, as seen in transport from the normal metal to the superconductor. 

A different class of qubits are potentially decoherence-free topological qubits. Conventional superconducting qubits such as superconducting flux, phase or transmon qubits have been implemented and coherence times about $100\mu \rm{sec}$ achieved by optimal control of their electromagnetic environment and intrinsic sources of decoherence.
The alternative qubit approach is, for example, based on the fermionic state formed by a combination of two Majorana particles. Due to the intrinsic topological protection from decoherence, the need for error correction is expected to be minimal \cite{Read_12}.

\newpage
\section{Quantum information circuits}
\begin{figure}[b!]
\begin{center}
\includegraphics[width=5cm]{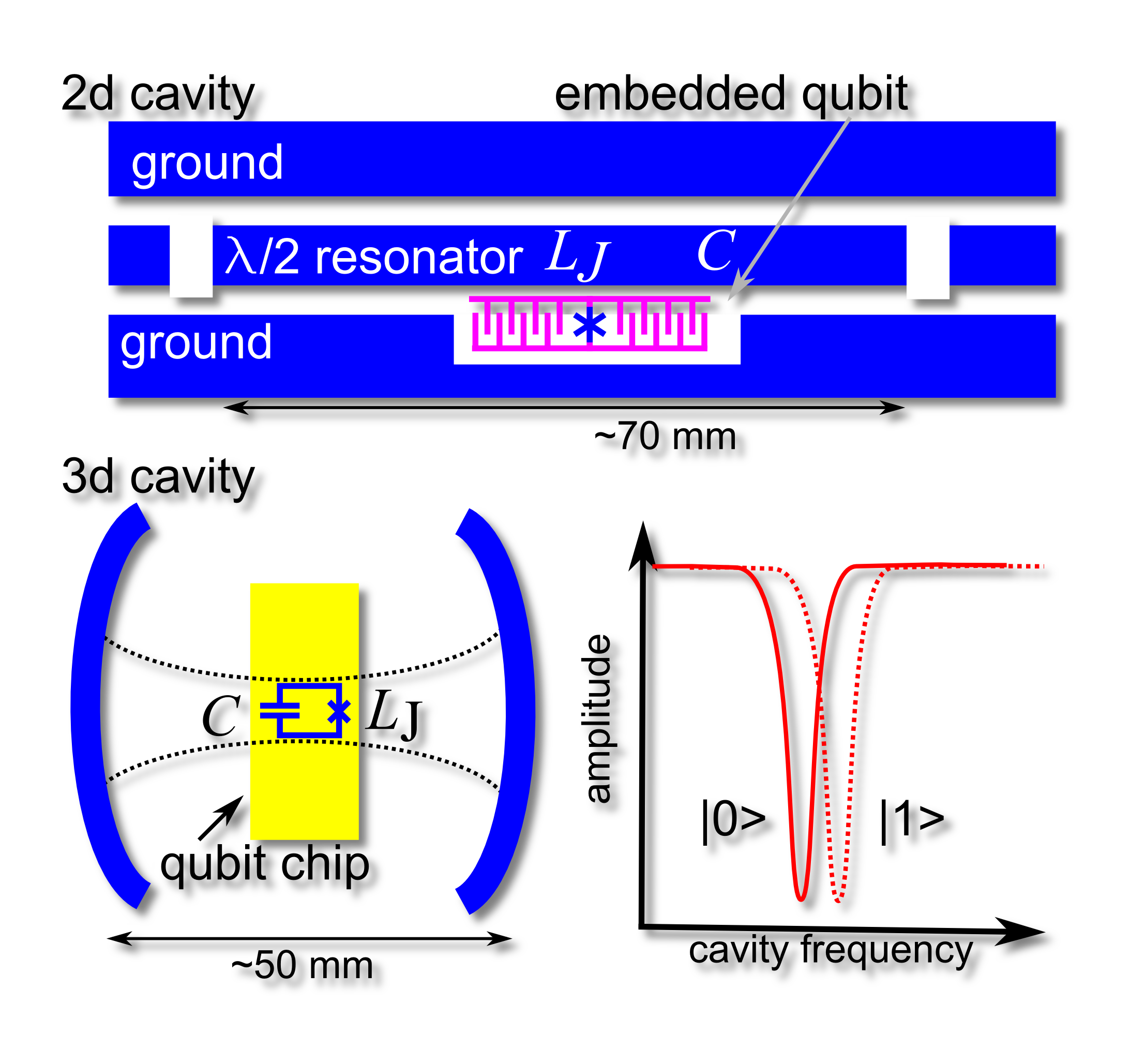}
\caption{\label{QuantumCircuit} Schematic of a qubit (here a transmon qubit) inside a 2d ({\bf top}) or 3d ({\bf bottom left}) cavity. Dielectric loss is reduced by better materials (2d implementation) or less surface dielectric contribution (3d implementation). The quantized cavity states suppressed the qubit coupling to the electromagnetic continuum giving rise to spontaneous emission. {\bf Bottom right:} The cavity response depends on the state of the dispersively coupled qubit ($|0\rangle$, $|1\rangle$) and is used for the qubit readout. While the 2d and 3d cavities have similar dimensions the 2d cavity can be meandered to reduce its footprint significantly. The resonator/qubit dimensions are not to scale. }
\end{center}
\end{figure}

Superconducting quantum information circuits are promising solid state candidates as building elements for quantum computers due to ultra-low dissipation, inherent to superconductors. The key element, the superconducting qubit, exists in three basic types as charge \cite{nakamuraNature99}, flux \cite{Mooij:1999:JosPersistCurrentQubit} and phase qubit \cite{Martinis:2002:RabiLJJ}. All of these first generation qubits may fulfill the five DiVincenzo criteria \cite{DiVincenzoCriteria} for a scalable quantum computer but are limited by low coherence. Depending on the Josephson over charging energy ratio $E_\mathrm{J}/E_\mathrm{c}$ (with $E_\mathrm{J}=\Phi_0 I_c/2 \pi$ and $E_\mathrm{c}=e^2/2C$) their coherence is affected by either charge noise (ultra-small capacitors, $E_\mathrm{c} \gg E_\mathrm{J}$), or dielectric loss (large capacitors, $E_\mathrm{c} \ll E_\mathrm{J}$). The second generation of superconducting qubit concepts (developed between 2002 to 2010) are based on hybrid qubits to minimize these environmental influences. For instance, charge-flux \cite{Vion:2002:ManipQuStatOfElCirc}, low impedance flux \cite{Steffen_PRL2010}, or flux-noise \cite{Zorin_PRB09} and charge-noise insensitive \cite{Koch_TransmonPRA07} qubits have been realized. For the third qubit generation, currently under development, the residual loss is being pushed below the threshold for quantum error correction. Advancement strategies involve scaling the qubit dimensions up to reduce the inevitable surface loss participation and employing better material with less dielectric surface states.

In fifteen years, impressive progress has been made to address, control, readout, and scale superconducting qubits, resulting, for example, in the proof of the violation of Bell's inequality, measurements of three qubit entanglement, quantum non-demolition readout, creation of arbitrary photon states, and circuit quantum electrodynamics in strong and ultra-strong coupling regimes.\\
In general qubits are limited by intrinsic decoherence processes which return an excited state to the ground state. In the Bloch picture of a two-level state, the energy relaxation time $T_1$ describes the longitudinal and the dephasing time $T_2$ the transverse relaxation. The physical origin for decoherence is attributed to, among others, coupling to discrete or continuum electronic defect states, flux or charge noise, or pair-breaking radiation.

The building blocks of quantum circuits are quantum gates operating on a small number of qubits. The quantum gate error budget has contributions from decoherence processes occurring during the gate operation such as signal noise or off-resonant excitations. Quantum error correction theory predicts that once the error rate of individual quantum gates is below a certain threshold, all errors can be corrected by concatenated quantum codes. These are based on stabilizer codes employing additional qubits (called ancilla qubits) coupled to the qubit of interest and thereby forming highly entangled, encoded states to correct for local noisy errors. By encoding one logical qubit in several physical qubits one can correct for bit flip or/and sign flip errors in the single logical qubit. As of 2013, the error correction threshold is achieved by the first superconducting quantum bits.

\subsection{Material and design considerations}
Superconducting qubits consist of linear (capacitor, inductor) and non-linear (Josephson junction inductance) circuit elements. One limiting coherence factor is the employed thin film material. In a resonant circuit the energy stored is $E_{\rm{total}}$, where on average the capacitive energy equals the inductive energy. The lifetime $T_1=1/\delta_\textrm{m} 2\pi f_q$ at the qubit frequency $f_q$ is a measure of the relative energy loss $\delta_{\textrm{m}}=\Delta E/E_{\rm{total}}$ per cycle time defined as
\begin{equation}
\delta_{\textrm{m}}=\frac{\Delta E}{2\pi E_{\rm{total}}}=\frac{\Delta E_C+\Delta E_L}{2\pi E_{\rm{total}}}=\delta_C+\delta_L\;.
\label{eqn:MeasuredLoss}
\end{equation}
Loss can be separated into contributions from the capacitive ($\delta_C$) and inductive ($\delta_L$) circuit elements. Superconducting qubits have been optimized by identifying the local contributions of capacitive and inductive losses to the linear and non-linear elements. The linear elements have been directly improved by considerable research activities on microwave kinetic inductance detectors \cite{DayNature03}. Progress in material research \cite{Megrant_APL12,BarendsAPL10,Vissers_APL10} and resonator surface \cite{Sandberg_TiN} led to $\Al$, $\Re$, $\Ti\N$ being the lowest-loss thin film superconducting resonators. Sapphire and undoped (high resistivity) silicon wafers are used as low-loss substrates.

The non-linear Josephson inductance, providing the qubit's anharmonicity, is conventionally formed by oxygen diffused, amorphous and defect-rich $\Al\O_x$ tunnel junctions.  A major contributor to relaxation processes in superconducting qubits originates from dielectric (i.e. capacitive) loss caused by microscopic two-level systems (TLS) in amorphous layers such as the surface oxide or tunnel barrier. TLS are commonly attributed to atoms or groups of atoms tunneling between two configuration states. The resonant response of an ensemble of TLS in a microwave field is found to decrease the coherence due to microwave absorption.
A small volume of the tunnel barrier or surface oxide layers implies that most operation frequencies do not put the qubit on resonance with these two-level systems. The tunnel junction loss contribution can be statistically avoided by scaling the non-linear $\Al\O_x$ tunnel junction down to tenths of microns dimensions. The surface loss arising from coupling to native oxides of a few nanometer thickness at the vacuum or substrate interfaces is suppressed by scaling the circuity up to a millimeter size and thereby reducing the electric field across, and the coupling to defect states within these oxides.

For example, the 3d transmon \cite{Paik_3d_11} uses a shunting capacitance $C$ with negligible surface loss due to the large electrode separation and ultra-small tunnel junction areas, see Fig. \ref{QuantumCircuit}. By placing the qubit chip in a machined microwave cavity of millimeter dimension the quantized mode spectrum protects the qubit from radiative electromagnetic loss and the bulk metal enclosure screens the pair-breaking blackbody radiation from higher temperature surfaces. 3d qubits were shown to have $T_1$ times above $100\;\rm{\mu sec}$. While their large volume per qubit may limit scalability beyond a certain number of qubits today's few-qubit experiments on quantum gates or quantum algorithms can be implemented directly and tested. Good scalability is expected from 2d qubit implementations as the resonator with the embedded qubit can be wrapped on the chip surface to reduce the footprint significantly. An increase in coherence is achieved by improved materials for capacitors, inductors and Josephson junctions. For example capacitors with electrodes having less interface oxides were shown to raise the qubit lifetime to the same order achieved by the 3d qubit. However, the absence of a quantized mode spectrum in the third dimension requires a well elaborated circuit geometry to tailor the electromagnetic environment and reduce coupling to a continuum of states.

\section{Metamaterials at microwave frequencies}
\begin{figure}[b!]
\begin{center}
\includegraphics[width=5cm]{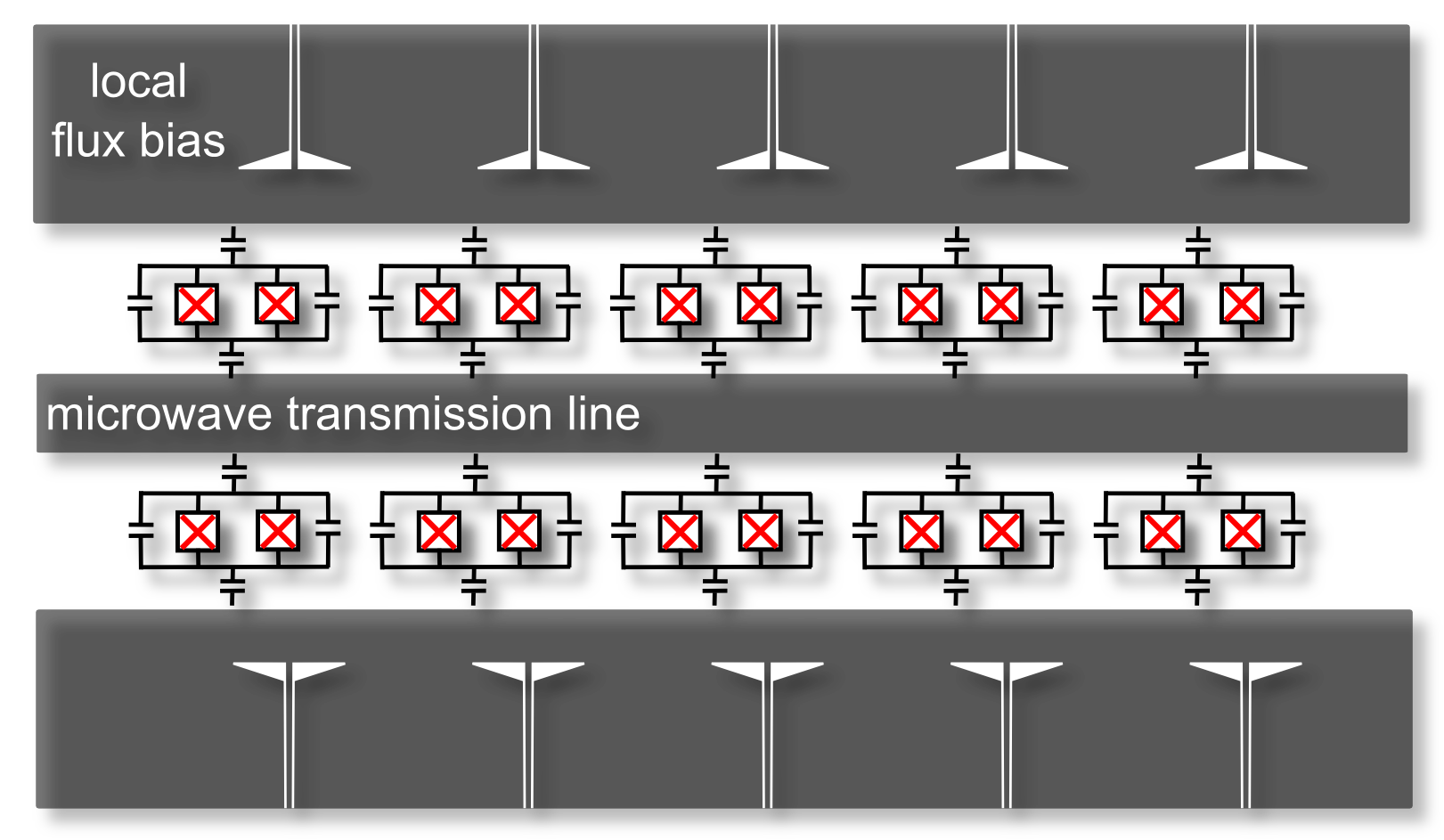}
\caption{\label{MM} Sketch of a chain of DC SQUIDs or qubits  placed at evenly spaced intervals and uniformly coupled a coplanar waveguide. Each element has an individual flux bias for selective control of its transition frequency. In general, the wave front propagation in the transmission line may be reversed, and contrary to the direction of energy flow if achieving 'left-handed' conditions. Frequency tuning of the band gaps is achieved inductively via a DC current in the center conductor of the waveguide or magnetic coils (on-chip or external).}
\end{center}
\end{figure}
Metamaterials are artificially engineered structures with electromagnetic properties that are not found in nature. The most striking example of such properties is a negative index of refraction, analyzed first in a theoretical work in 1967 \cite{Veselago_67} and experimentally realized in 2001 in arrays of split-ring resonators  and rods \cite{Pendry_99,Shelby_Science01}. Negative real parts of both dielectric permittivity $\epsilon$ and magnetic permeability $\mu$ can be realized by working at frequencies near a resonance of an artificial material electromagnetic resonator. For example in metals to the plasma resonance, the permittivity is negative below and  positive above the resonance. Below the resonance frequency, the waves oscillate slowly enough for the electrons to follow. Above the resonance frequency, the inertia of the electrons prevents the electrons from oscillating in proper phase with the incident wave. The magnetic permeability, in contrary, becomes negative above some -in general-  different frequency. In a double negative index material, the wave front travels toward the source whereas the energy propagation (expressed by the Poynting vector $\vec{S}$) is transmitted away from the source (as in conventional materials). Such materials are often referred to as being 'left handed' as the electric and magnetic fields, and the direction of propagation of the electromagnetic wave obey the left-hand rule. A lossless negative-index material would, in principle, allow an imaging resolution well below the diffraction limited by focusing the entire spectrum containing both the propagating as well as the evanescent spectra.

However, in conventional metallic nanostructures the resonant current enhancement causes significant losses limiting the field amplification. And the resonance frequency is determined by the geometry, thus complicating the frequency tunability. Using superconducting resonators with Josephson junctions as tunable, non-linear inductors allows improving the performance by adding high field enhancement over a broad, tunable microwave range.

\subsection{Classical metamaterials}

The superconducting version of the conventional split-ring resonator based metamaterial is a two-dimensional array of thin film resonators. Such an array exhibits large negative magnetic response above the resonance frequency and, if including SQUIDs, the frequency can be tuned by an external perpendicular magnetic field. Experiments with superconducting niobium split-ring resonators showed evidence of negative effective permittivity, permeability, and a negative effective index pass-band in the superconducting state at GHz frequencies \cite{Ricci_APL05}. A nonlinear left-handed transmission line can be formed by Josephson junctions as shunt inductances. For instance, a one-dimensional superconducting metamaterial, tunable over a broad frequency band, was realized by coupling a microwave waveguide to an array of 54 RF-SQUIDs. Using the nonlinear inductance of it's basic building block the resonance frequency is tunable in-situ by applying a DC magnetic field \cite{Butz_2013}. The effective magnetic permeability of this artificial material is determined from the complex scattering matrix. Such one-dimensional Josephson junction arrays (see sketch in Fig \ref{MM} for DC SQUID chain) with generalized unit cells allow to engineer the band gaps in the electromagnetic spectrum in full analogy to electronic band gaps in crystals. Frequency-tunable right- and left-handed linear and nonlinear dispersion relation element can be merged for novel and innovative conceptual advance of microwave elements and devices.\\
A different kind of tunable metamaterial was realized in a SQUID chain operating in the lower GHz frequency band \cite{Castellanos_NatPhys08}. This general-purpose parametric device is capable of squeezing the quantum noise of the electromagnetic vacuum. Squeezed states, having less uncertainty in one observable than the vacuum state are used for enhanced precision measurements. For instance amplitude-squeezed light improves the readout of very weak spectroscopic signals, while phase-squeezed light improves the phase readout of an interferometer. The parametric oscillator is used for low-noise amplifiers, for example for quantum-state readout in the microwave domain and quantum non-demolition measurements.

\subsection{Quantum metamaterials}

Besides the above-mentioned promising improvements of known metamaterial functionalities through reduced losses and frequency tunability, superconducting metamaterials offer a major novel and potentially disruptive properties that originate from their quantum nature.
The quantum electromagnetic properties of such superconducting qubits can be controlled and tuned by on-chip or external DC magnetic fields. The experimental realization of controlled coupling between the quantum electromagnetic field and chains/arrays of qubits that evolve quantum-coherently with the field may open practical routes towards the very exciting field of quantum metamaterials. They can be realized, for instance, through tailored chains of qubits that are embedded into superconducting transmission-line resonators. This opens entirely novel opportunities such as the realization of lasing with very few (or even just one \cite{AstafievNature07}) qubits.\\
First thoughts of using macroscopic superconducting quantum metamaterial have been suggested in theoretical works. For the quantum version of SQUID arrays as a left-handed metamaterial the quantum expressions for the associated negative refractive index were derived in Ref. \cite{Du_PRB06}. The propagation of a classical electromagnetic wave through a transmission line formed by superconducting qubits embedded in a superconducting resonator was considered in Ref. \cite{Rakhmanov_PRB08}. In particular, the spectroscopic properties of such a quantum Josephson transmission line will be controlled by the quantum coherent state of the qubits. And a chain of qubits coupled the transmission line can modify the band-gap structure to slow and stop the microwave propagation \cite{Zhou_08}.

A single qubit coupled to a harmonic oscillator is described by the Jaynes-Cummings \cite{JaynesCummings_63,Blais_04} model, and it's generalization to many qubits not interacting with each other is given by the Tavis-Cummings model \cite{TavisCummings_68}. This model provides the framework to study collective properties of the qubits for the coherent transmission of microwave photons through a microwave resonator coupled to tunable qubits periodically placed in the gap and capacitively coupled to the resonator. The Dicke model \cite{Dicke_54} describes the coupling between such a collection of two-level systems and a single photon mode. For increasing light-matter coupling the model predicts excitons giving rise to many cooperative radiation phenomena such as superradiance and super-fluorescence \cite{BonifacioPRA_75}. The counterpart of the collective excited state radiative decay is a destructive interference effect, termed subradiance, leading to the partial trapping of light in the system. Adding nearest-neighbor interaction in linear chains of Josephson qubits leads to interesting dynamical entanglement properties of coupled many-body systems \cite{Tsomokos_NJP07}. Multiple superconducting qubits can coupled identically to the field mode of the same cavity, for example by optimizing the placement of qubits at field antinodes of a distributed resonator \cite{Delanty_11}. For large couplings both superradiant microwave pulses and phase multistability are expected. In the case of non-uniform coupling rates subradiant transitions are induced \cite{Ian_PRA12}. The resulting excitation spectrum depends on the amount the inhomogeneous system deviates from the homogeneous case, and crosses from Frenkel- to Wannier-type (i.e. localized to non-localized) excitons for increasing larger distances in the superconducting qubit chain.\\
From an experimental point of view, the coherent quantum dynamics of superconducting qubits coupled to the electromagnetic field in a transmission line opens novel opportunities for material-induced coherent transformation (e.g. light squeezing, coherent down- and up-conversion, etc.) of incident radiation. So far mostly experiments involving one qubit have been performed. Resonance fluorescence \cite{Astafiev_Science10} and electromagnetically induced transparency \cite{Abdumalikov_PRL10} on an individual qubit were observed on one qubit coupled to a transmission line. Coherent population trapping was found in a phase qubit \cite{Kelly_PRL10}. And an electromagnetically induced transparency qubit was use to realize a single-photon router, where an incoming signal can be routed to the output port \cite{Hoi_PRL11}.\\

\section{Quantum phase slip}

\subsection{Basic concept}

\begin{figure}[h!]
\begin{center}
\includegraphics[width=0.5\textwidth]{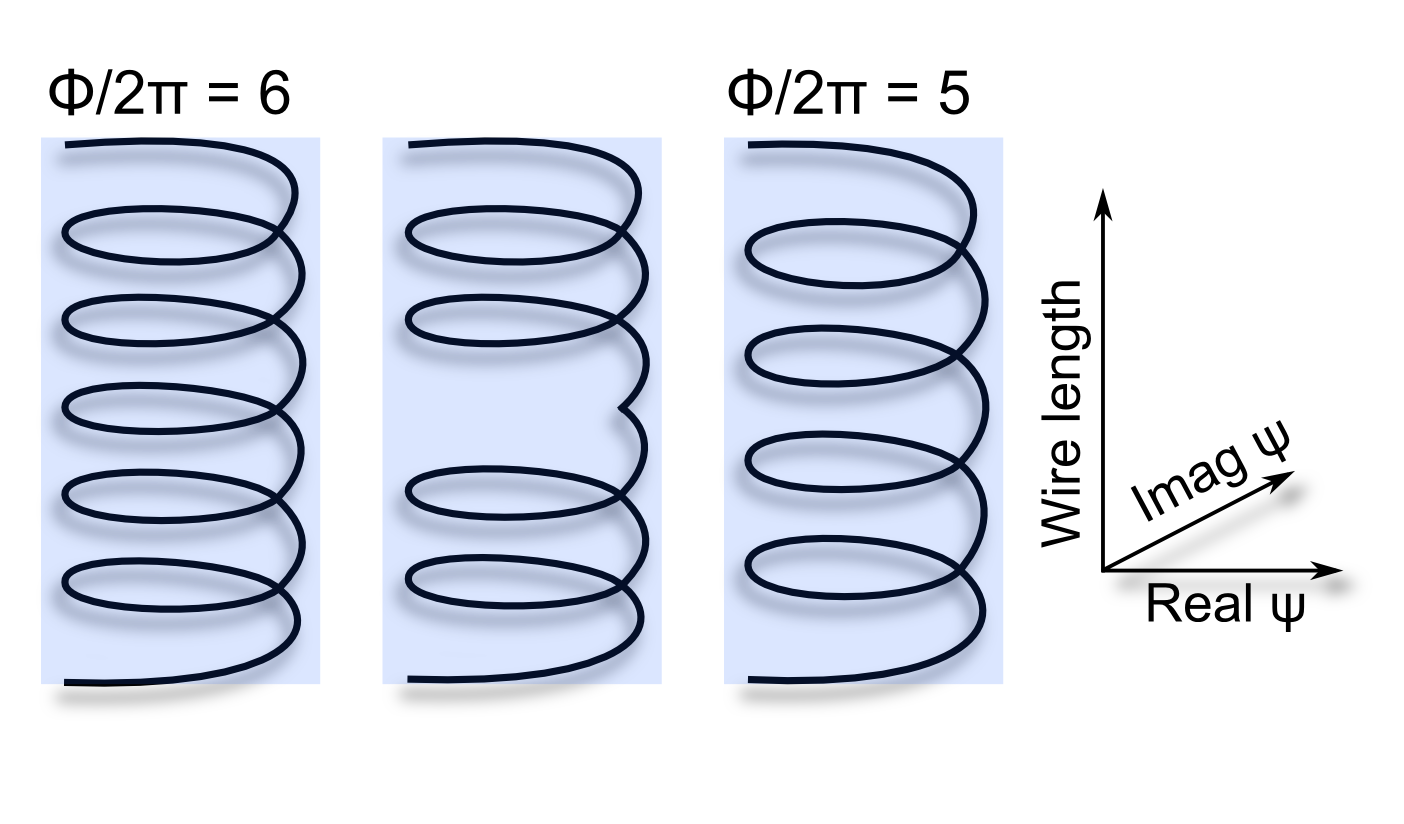}
\caption{\label{PSscheme} Ilustration of a phase slip process. In a narrow superconducting wire the accumulated phase $\phi$ of the order parameter $\psi$ can slip by $2\pi$ if the width is comparable to the coherence length.}
\end{center}
\end{figure}

In a quantum phase slip circuit, a one dimensional superconducting nanowire is used as a non-linear element connecting two conventional superconductors. The research on phase slips has its origins in the late 1960s, where transport measurements observed a finite resistance of superconducting wires in the close vicinity to the superconducting transition temperature.
In such a wire, phase slips of the superconducting order parameter $\psi$ can occur if the width of the wire is comparable to the superconducting coherence length $\xi$. This is easy to fulfill even for moderate wire width ($\sim \mu$m) close to the superconducting transition temperature, where $\xi$ diverges.\\
An applied voltage $V$ is increasing the phase difference $\phi$ between the ends of the wire terminals  according to d$\phi/$d$t = 2e/\hbar\,V$. Here, $e$ is the elementary charge and $\hbar$ the reduced Planck constant.  Without a phase slip, the number of phase turns increases until the current flow $I\sim\nabla\phi(\vec{r})$ reaches a critical value and superconductivity is lost.\\
In a phase slip capable superconducting wire, $\psi$ can vanish for a short period of time over the length scale of $\xi$ and allows to reduce the acquired phase by a $2\pi$ 'slip'. Since the phase around an elementary superconducting flux vortex is also 2$\pi$, a phase slip is equivalent to a flux vortex passing across a two or three dimensional wire.\\
In an energy potential picture, the two states of the wire, before and after the phase slip, have the same energy level. However, a phase slip event has to overcome the the condensation energy barrier of the superconductor. If the available energy is large enough for a thermally activated phase slip, the phase is allowed to change by passing over the condensation barrier potential.
These phase slips happen as a stochastical process and are therefore dissipative. The energy transferred per phase slip is $I\Phi_0$, where $\Phi_0 = h/2e$ is the superconducting flux quantum and $I$ the current passing through the wire.\\
In 1988, Giordano \cite{Gi88} observed a non-vanishing electrical resistance with very small ($\sim$500~nm) $\In$ wires well below temperatures that would allow for a thermal activation of phase slips. Although the experimental conditions led to some controversies, a quantum mechanical tunneling processes was discussed to explain the data. The main idea pointed out in the publication was that the phase may not change in the low temperature regime by passing over the condensation energy barrier, but instead by quantum mechanical tunneling through it. This is in analogy to the well known macroscopic quantum tunneling process (MQT).\\
In the following years more experiments on superconducting nanowires have been carried out. Noticeably the Tinkham group (e.g. \cite{Be00,La01}) studied many samples. They used ultra narrow bridges of free hanging carbon nanotubes as a template for superconducting $\Mo\Ge$ wires. Although these wires had diameters down to a few nanometer, the results could not unambiguously prove the existence of a quantum mechanical effect, still strong evidence for \emph{quantum phase slip}, QPS, was provided.
In-depth analysis of the experiments and extensive theoretical efforts led to models describing the QPS process on a microscopic as well as on a macroscopic level (for a recent review, also on thermally activated phase slips, see e.g. \cite{Ar08}).\\
Briefly summarized the following is known about nanowires exhibiting the QPS effect. The QPS coupling energy $E_\mathrm{s}$ depends linearly on the length $L$ of the wire in units of $\xi$, and exponentially on the normal state resistance R$_n/R_q\times\xi/L$ over a wire length $\xi$. Here $R_n$ is the wire's sheet resistance and $R_q = h/(2e)=$ 6.45 kOhm is the superconducting resistance quantum. Advantageous to get to the QPS regimes is therefore a material with a long superconducting coherence length $\xi$ and a high normal state sheet resistance. Given this, wires with a smaller diameter will allow for a higher $E_\mathrm{s}$ and therefore higher phase-slip rate.\\
QPS wires have been fabricate down to a few nanometer employing different techniques and a wide range of superconducting materials. One experimental example is given below in section \ref{phaseslipqubit}

A theoretical paper by Mooij and Nazarov  \cite{Mo06} drew a lot of attention in 2006. The authors expanded the field by deriving a fundamental duality between the Josephson and the quantum phase slip effect.  Two extreme cases for both junction types are discussed:
\begin{itemize}
\item The Josephson junction in the superconducting state, namely the Josephson coupling energy $E_\mathrm J$ is much larger than the charging energy $E_\mathrm{c}=e^2/2C$ of the junction, $C$ is the junction's capacitance. The junction exhibits for a supercurrent up to a critical \emph{current} $I_c$ without a voltage drop. This is the coherent, dissipation free regime where Josephson junctions down to a size of $\sim 1\mu$m$^2$ are classically operated.
\item Correspondingly, with the QPS energy $E_\mathrm{s}$ large compared to the inductive energy $E_\mathrm{L}=\phi_0^2/2L$ the QPS junction is in a strong quantum phase slip regime with no current flow up to a critical \emph{voltage} $V_c$. $L$ is the total inductance of the wire. The phase coherence along the wire is very low and therefore charge transport is suppressed.
\item In the opposite Josephson junction case, $E_\mathrm{c} > E_\mathrm{J}$ and embedded in a proper environment, the system is insulating and a current flow is only admitted with the charging energy $E_\mathrm{c}$  provided above a critical \emph{voltage} $V_c$. This is the well known regime of ultra-small Josephson junctions and arrays of ultra-small Josephson junctions with a coulomb blockade of current.
\item If the QPS junctions inductive energy $E_\mathrm{L}$ is larger than $E_\mathrm{s}$, there exists  a superconducting phase coherence along the wire and a supercurrent is observable. Since the quantum phase slip energy $E_\mathrm{s}$ is non-zero, quantum phase slips happen at a lower rate compared to the strong phase slip regime.
\end{itemize}
It is interesting to note that the duality of JJ and QPSJ is exact with respect to the exchange of the canonically conjugated quantum variables, phase and charge. Also the existence of a kinetic capacitance in duality to the kinetic inductance of the nanowires was predicted for QPS wires.\\
As a consequence of the duality of Josephson junctions and QPS junctions and previous theoretical studies \cite{Za97,Go01,Bu04} the question was raised, if it is possible to observe a dissipation free \emph{coherent} quantum phase slip process. This is addressed in the next section.

\subsection{\label{phaseslipqubit}Phase slip flux qubit}

\begin{figure}[t!]
\begin{center}
\includegraphics[width=0.9\textwidth]{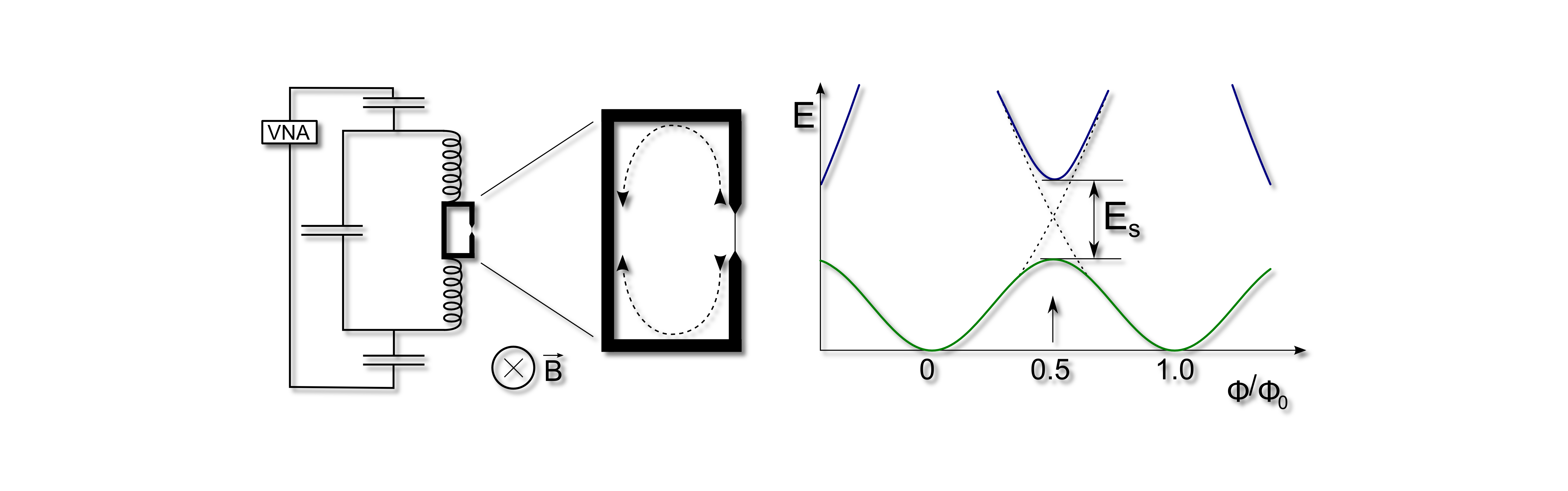}
\caption{\label{PSQubit} {\bf Left:} Phase slip flux qubit embedded in a superconducting microwave resonator. {\bf Right:} Energy spectrum versus applied flux. The energy of the loop without a QPS capable wire is given by the dashed black line. Quantum phase slip coupling of two neighboring flux states lifts the degeneracy at 0.5$\Phi_0$ by $E_\mathrm{s}$.}
\end{center}
\end{figure}
Built upon a previous work by Mooij and Harmans \cite{Mo05}, an international collaboration around Astafiev and coworkers presented in a recent publication \cite{As12} the exact duality outlined above by exploring a phase-slip flux qubit.\\The phase slip qubit consists of a superconducting loop with an embedded QPS wire, see Fig.~\ref{PSQubit}, left. In the superconducting state the flux in the loop is fixed and can only change by a phase slip event.
The qubit is very similar to the well known Josephson junction based flux qubit.
In case of a vanishing $E_\mathrm{s}$, the energy spectrum of the loop is given by $E=(\Phi_\mathrm{ext}-N\Phi_0)^2/2L$, where $L$ is the inductance and $N$ the number of flux quanta in the loop, depicted as dashed parabolas in the energy spectrum Fig.~\ref{PSQubit}.\\If an external flux $\Phi/\Phi_0=0.5$ is applied to the qubit with a non-vanishing phase slip energy $E_\mathrm{s}$, a flux vortex is allowed to periodically enter (or leave) the superconducting loop. At 0.5$\Phi_0$ the neighboring flux state degeneracy is lifted and a gap for the two relevant qubit states separated by $E_\mathrm{s}$ opens in the spectrum.\\
Astafiev \emph{et al.} showed exactly this behavior by strongly coupling a QPS-qubit to a superconducting resonator (Fig.~\ref{PSQubit}, left sketch) and measuring the resonator transmission spectrum depending on applied magnetic flux.
They demonstrated that quantum phase slips can be induced by flux bias driving, thus non-stochastical \emph{coherent quantum phase slips} are possible. For the first time it was shown experimentally,  that there can be a coherent coupling of flux states across a QPS junction in exact duality to the coherent coupling of charge states across a Josephson junction. Furthermore, $E_\mathrm{s}$ was for the first time directly measured, confirming theoretical models for QPS in the weak phase slip regime.  Astafiev \emph{et al.} used a  QPS qubit where loop and wire were made from a highly resistive, amorphous $\In\O$ film with a superconducting transition temperature of 2.7~K. The wire had a length of 400~nm and width of 40~nm, defined by e-beam lithography. $E_\mathrm{s}$ was measured to about 5~GHz with an $\xi$ estimate in the range of 10-30~nm. All these parameters are accessible by conventional fabrication techniques, encouraging for more QPS related experiments.

\subsection{Constant current steps}

\begin{figure}[t!]
\begin{center}
\includegraphics[width=0.4\textwidth]{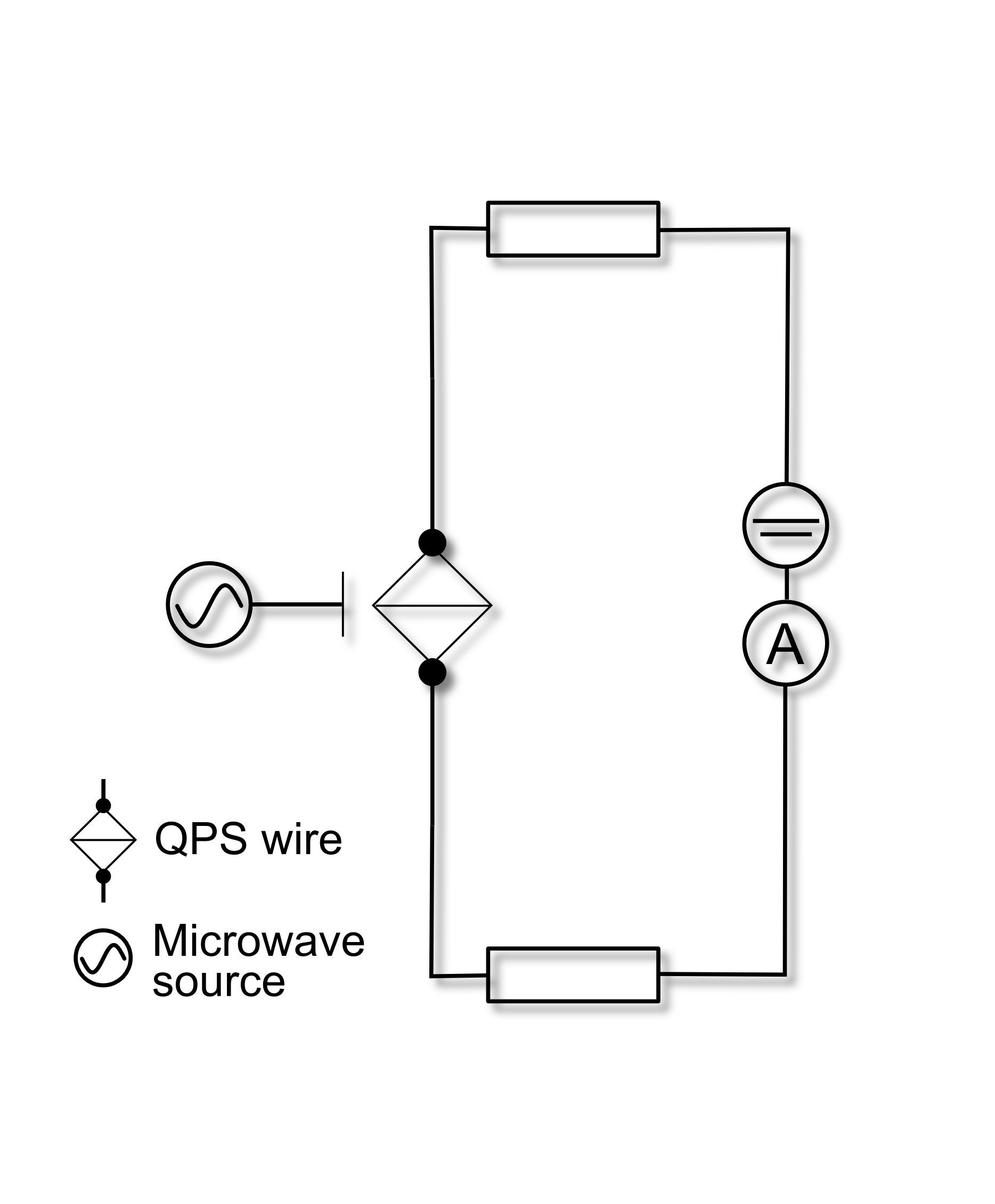}
\hspace{5mm}
\includegraphics[width=0.38\textwidth]{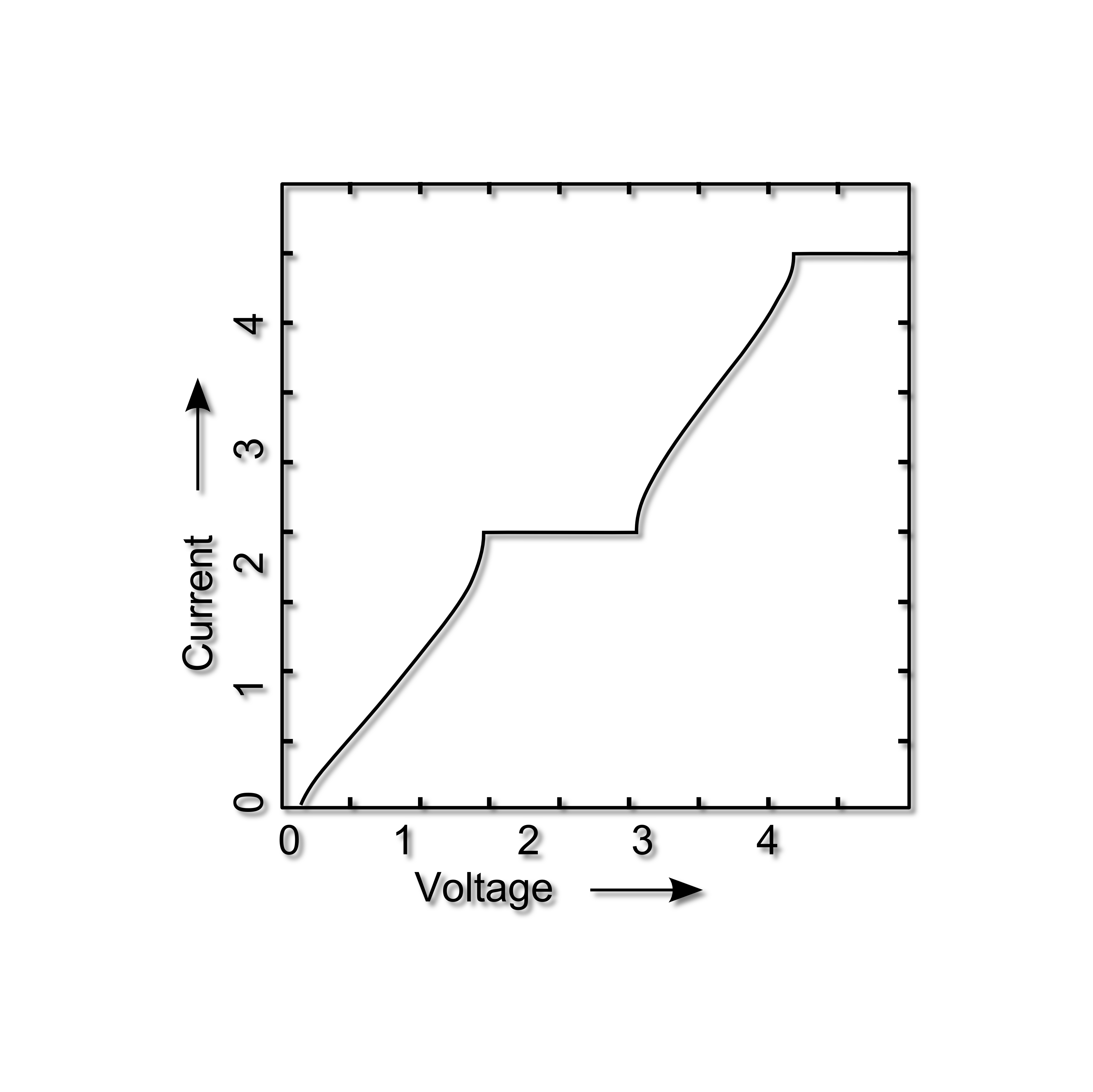}
\caption{\label{QPS_current_standard} {\bf Left:} Simplified schematics of a current  circuit based on a QPS wire. {\bf Right:} IV characteristics of a QPS wire under microwave radiation.}
\end{center}
\end{figure}
Assuming the duality between the Josephson junction and the phase slip junction outlined above is correct, there must be an equivalent to the constant voltage steps found in the $I-V$ characteristics of the Josephson junction under microwave radiation (Shapiro steps). If found, they might be of scientific and technological interest, due to their potential for being used as a current standard.\\
The QPS junction in the strong phase slip regime embedded in a dissipative environment, see Fig.~\ref{QPS_current_standard}, left, can be described by a semiclassical RCSJ type circuit model with $I$ and $\phi$ replaced by their canonical conjugated variables $V$ and $q$. The voltage drop over a QPS junction is then given by
\[ V(t)  = V_c \sin( 2\pi q)  + 2e \left( L \frac{\mathrm{d}^2q}{\mathrm{d}t^2} + R \frac{\mathrm{d}q}{\mathrm{d}t}\right)  \]
The model is of the well known tilted washboard potential type, with the tilt $V$ and the position of the virtual  particle $q$  \cite{Mo06}. If an oscillating voltage with a frequency $\nu$, phase locked on the QPS rate, is applied to a QPS junction, plateaus of constant current develop in the $I-V$ characteristic at a current $I=n\, 2e\,\nu$, where $n= 1,2,3,...$ is the step, see Fig.~\ref{QPS_current_standard}, right plot.\\
First experiments looking for current steps in QPS junctions under microwave radiation have been carried out, but were not successfully observing the steps, see e.g. Refs. \cite{Ho12,We13}. In the ultra-low temperature experiments ($T<$100~mK), the heat produced by the high ohmic $\Ni\Cr$  on-chip bias resisters was found to be significant. This prevents the QPS wire to be intrinsically cold enough and the conductance not being dominated by parasitic quasiparticle channels.\\
The existence of a fundamental and direct current-frequency relation utilizing QPS wires remains therefore an open experimental question.

The authors acknowledge discussions with S. Butz and P. Jung.

\end{document}